\documentclass{aa}  

\usepackage{graphicx}

\usepackage{booktabs}
\usepackage{multirow}
\usepackage{txfonts}
\usepackage[colorlinks=true, citecolor=blue,
linkcolor=blue, urlcolor=blue]{hyperref}
\usepackage{epstopdf}
\usepackage{xcolor}
\newcommand{\hl}{\color{black}}
\usepackage{amstext}
\begin{document}

   \title{Improving the $\Gamma$-functions Method for Vortex Identification}

   \author{Quan Xie
          \inst{1,2}
          \and
          Jiajia Liu
          \inst{1,2,*}
          \and
          Robert Erd\'elyi
          \inst{3,4,5}
          \and
          Yuming Wang
          \inst{1,2}
          }

   \institute{National Key Laboratory of Deep Space Exploration, School of Earth and Space Sciences, University of Science and Technology of China, Hefei 230026, China
              \email{jiajialiu@ustc.edu.cn}
              \and
              CAS Center for Excellence in Comparative Planetology/CAS Key Laboratory of Geospace Environment/Mengcheng National Geophysical Observatory, University of Science and Technology of China, Hefei, 230026, China
              \and
              Solar Physics and Space Plasma Research Centre (SP2RC), School of Mathematical and Physical Sciences, University of Sheffield, Sheffield S3 7RH, UK
              \and
              Department of Astronomy, Eötvös Loránd University, Budapest, Pázmány P. sétány 1/A, H-1117, Hungary
              \and
              Gyula Bay Zoltan Solar Observatory (GSO), Hungarian Solar Physics Foundation (HSPF), Pet\H{o}fi t\'er 3., Gyula H-5700, Hungary
             }

   \date{Received XXXX; accepted XXXX}

 
  \abstract
   {Vortices have been observed at various heights within the solar atmosphere and are suggested to potentially play great roles in heating the solar upper atmosphere. Multiple automated vortex detection methods have been developed and applied to detect vortices.}
   {We aim to improve the \(\Gamma\)-functions method for vortex identification by optimizing the value of \(\Gamma_{1min}
\) and the approach to calculate \(\Gamma_1\) and \(\Gamma_2\) used to determine vortex center and edge. In this way, we can detect vortices more accurately and enable more statistical studies that can improve our knowledge of the generation and evolution of vortices in the solar atmosphere.}
   {We apply the automated swirl detection algorithm (ASDA, one representative of \(\Gamma\)-functions method) with different parameters to various synthetic data, with each containing 1000 Lamb-Oseen vortices, and search for the optimal \(\Gamma_{1min}\) and kernel size when calculating $\Gamma_1$ and $\Gamma_2$. We also compare another detection method using simulation and observational data to validate the results obtained from the synthetic data.}
   {The best performance is found with the Optimized ASDA, which combines different kernel sizes {\hl (5, 7, 9, and 11)} to calculate \(\Gamma_1\) and \(\Gamma_2\) with a fixed \(\Gamma_{1min}\) = 0.63 to detect vortex center. We find that more vortices can be detected by the Optimized ASDA with higher location, radius, and rotation speed accuracies. The above results are further confirmed by comparing vortices detected by the Optimized ASDA and the SWIRL method on CO5BOLD numerical simulation data and SST observational data.}
  {}


   \maketitle
%

\section{Introduction}
    
Rotational motions, spanning a wide range of spatial scales, have been widely observed at various heights within the solar atmosphere\citep[e.g.,][]{wang1995vorticity, li2012solar, liu2012slow, su2012solar, wedemeyer2012magnetic, panesar2013solar, wang2016tornado, liu2019automated, liu2019evidence, tziotziou2023vortex}. Numerous studies have highlighted their potentially significant role in channeling energy to the upper solar atmosphere. It is widely accepted that various modes of MHD waves, especially Alfv{\'e}n waves/pulses, could be associated with various vortices, as demonstrated by numerical simulations \citep[e.g.,][]{shelyag2013alfven, chmielewski2014numerical, mumford2015generation, mumford2015photospheric, liu2019evidence, battaglia2021alfvenic, kesri2024dependence}. Observational evidence also supports this connection. For instance, \cite{wedemeyer2012magnetic} reported that magnetic tornadoes act as energy channels into the solar corona, based on manual detection of chromospheric vortices. Additionally, \cite{liu2019evidence}, by detecting small-scale vortices using an automated algorithm, provided evidence that ubiquitous Alfv{\'e}n pulses, triggered by photospheric vortices, transport energy to the upper chromosphere. {\hl \cite{tziotziou2019persistent} performed spectral analysis of a 1.7-hour vortex flow characterized by multiple intermittent chromospheric swirls and found dominant oscillations around 4 minutes, with both swaying (200–220 s) and rotational motions, as well as significant oscillatory power up to 10 minutes, which indicates the presence of various MHD wave modes at different heights. \cite{tziotziou2020persistent} further provided observational evidence of fast kink and localized torsional waves, which are associated with small chromospheric swirls and swaying motions within a persistent vortex flow.} Small-scale vortices in the photosphere are also believed to contribute to energizing the upper atmosphere \citep[e.g.,][]{parker1983magnetic, velli1999alfven, shelyag2013alfven}. Furthermore, theoretical studies suggest that rotational motions could generate upward mass and momentum transfer, thereby leading to the generation of small-scale jets \citep[spicules, e.g.,][]{scalisi2021propagation, scalisi2023generation}.

Over time, vortex motions found in the solar atmosphere have been classified into several types based on their dynamic characteristics and formation mechanisms. The term "tornado" in solar context was first introduced by \cite{pettit1932characteristic} to describe vortex motions, particularly those associated with prominences. This category includes solar tornadoes \citep{pike1998rotating}, giant tornadoes \citep{li2012solar, su2012solar}, magnetic tornadoes \citep{wedemeyer2012magnetic}, and small-scale tornadoes \citep{tziotziou2018persistent}. Smaller vortical phenomena have been called ``swirls'', which include chromospheric swirls \citep{wedemeyer2009small}, small-scale swirls \citep{shetye2019multiwavelength}, magnetic swirls \citep[e.g.,][]{chmielewski2014numerical,murawski2018magnetic}, and downdraft swirls \citep{moll2011vortices}. Additionally, the term ``vortex'' is often used in theoretical contexts derived from simulations, such as vortex tubes \citep{muthsam2010antares}, horizontal vortex tubes \citep{steiner2010detection}, magnetized vortex tubes \citep{kitiashvili2013ubiquitous}, and kinetic \mbox{(K-)} and magnetic (M-) vortices \citep[e.g.,][]{silva2020solar, silva2021solar}. One notable exception is the ``photospheric intensity vortex'', which, unlike the others, originates from observational data and is often called ``swirls'' \citep[e.g.,][]{giagkiozis2018vortex, liu2019automated, liu2019evidence}. In this work, we focus on the automated detection of small-scale vortices (also named ``swirls'').

Detecting small-scale vortices accurately and efficiently from observations has long been a key challenge. The first step in identifying numerous small-scale vortices from observational images is to reconstruct the horizontal velocity field of each image. Techniques such as Local Correlation Tracking \citep
[LCT;][]{november1988precise}, Fourier Local Correlation Tracking \citep[FLCT;][]{fisher2008subsurface}, and Coherent Structure Tracking \citep[CST;][]{rieutord2007tracking} use two consecutive intensity images to estimate the velocity field at the photosphere. DeepVel \citep{ramos2017deepvel} and its U-Net version, DeepVelU \citep{tremblay2020inferring}, are deep fully convolutional neural networks that serve as end-to-end approaches for estimating the velocity field also from two consecutive images. \cite{tremblay2018reconstruction} compared DeepVel, LCT, FLCT, and CST, and found that FLCT performs adequately at subgranular and granular scales (although it is outperformed by DeepVel), but is the most effective at mesogranular and supergranular scales. DeepVel, however, could potentially outperform the other methods if trained with data at the corresponding spatial resolution. {\hl It is worth noting that LCT-based methods (e.g., LCT and FLCT) are not always reliable for reconstructing the horizontal velocity field, usually underestimating the actual speed \citep[e.g.,][]{verma2013evaluating, liu2019automated, liu2019evidence, xie2025photospheric, Liu2025suvel}. This point will also be discussed in detail later.} 

Based on the estimated velocity fields, due to the biases and limitations inherent in manual detection, various automated methods have been proposed. For example, \cite{strawn1999computer} introduced the Maximum Vorticity Method, which identifies overlapping vortex centers with the same sense of rotation when the overall velocity field outlines a single rotational center. Furthermore, \cite{jiang200514} developed an algorithm based on the Maximum Vorticity Method. Another widely adopted approach is the $\Gamma$-functions method, proposed by \cite{graftieaux2001combining}, which accurately identifies vortex centers and boundaries. This method has led to further automated algorithms, such as the Automated Swirl Detection Algorithm (ASDA) developed by \cite{liu2019automated} and the Advanced Gamma Method (AGM) proposed by \cite{yuan2023advanced}. Moreover, based on the velocity gradient tensor, the Rortex criterion was introduced by \cite{tian2018definitions} and \cite{liu2018rortex} to measure the strength of pure local rotation without contamination from shear. This makes Rortex a reliable quantity for inferring rotational flow properties. Building on this, \cite{cuissa2022innovative} developed the SWirl Identification by Rotation-centers Localization (SWIRL) algorithm, which applies the Rortex criterion for detecting swirls. 

Though the detection methods mentioned above have overcome certain limitations and achieved notable progress, they still exhibit some shortcomings. For instance, the  $\Gamma$-functions method identifies vortex centers and boundaries using the $\Gamma_{1}$ and $\Gamma_{2}$ criteria, respectively. \cite{graftieaux2001combining}
 proposed that regions where \( \lvert \Gamma_{2} \rvert > 2/\pi \) are predominantly governed by rotation, and points where \( \lvert \Gamma_{2} \rvert = 2/\pi \) are classified as the boundaries of vortices. However, for center identification, they merely suggested that \( \lvert \Gamma_{1} \rvert \) reaches values between 0.9 and 1 near the vortex center. A strict and universally accepted threshold for \(\Gamma_1\) (denoted as \(\Gamma_{1min}\) hereafter) to precisely identify vortex centers is still lacking. In practice, a point is considered the center of a vortex if \( \lvert \Gamma_{1} \rvert \) exceeds $\Gamma_{1min}$. For instance, \cite{liu2019automated} defined a point where \( \lvert \Gamma_{1} \rvert \geq 0.89 \) as a vortex center, whereas \cite{yuan2023advanced} used a lower threshold of 0.75 for $\Gamma_{1min}$. Additionally, the selection of kernel size ($ks$) used to calculate $\Gamma_1$ and $\Gamma_2$ is also unclear. \cite{liu2019automated} used a fixed kernel size of $ks$ = 7 in ASDA, while \cite{yuan2023advanced} proposed an adaptive method which is illustrated in detail in Sect. \ref{ks}. Which approach is more suitable and whether there is a more accurate method to calculate the $\Gamma$ functions still needs further exploration.

In this paper, we improve the $\Gamma$-functions method by searching an appropriate $\Gamma_{1min}$ and an optimal method to calculate $\Gamma_1$ and $\Gamma_2$. The paper is organized as follows. First, in Sect. \ref{method} we introduce the $\Gamma$-functions method and other methods utilized in the study. Sect. \ref{experi and res} describes the details of the experiments conducted and the corresponding results. We conclude our findings and make discussions in Sect. \ref{conclus and discuss}.

\section{Method} \label{method}

\subsection{\texorpdfstring{$\Gamma$}{}-functions method}
The main principles of the $\Gamma$-functions method are two functions $\Gamma_1$ and $\Gamma_2$, used to identify vortex centers and boundaries, respectively. \cite{graftieaux2001combining} defined these two functions as follows:
\begin{equation}
    \begin{aligned}
        &\Gamma_1(P)=\frac{1}{N}\sum_S\frac{({\hl \boldsymbol{PM} }\wedge {\hl \boldsymbol{U_M} })\cdot {\hl \boldsymbol{z} }}{||{\hl \boldsymbol{PM} }||\cdot||{\hl \boldsymbol{U_M} }||}=\frac{1}{N}\sum_S\sin(\theta_M),\\
        &\Gamma_2(P)=\frac{1}{N}\sum_S\frac{[{\hl \boldsymbol{PM} }\wedge{\hl \boldsymbol{U_M}}-{\hl \boldsymbol{\tilde{U}_P}}]\cdot {\hl \boldsymbol{z} }}{||{\hl \boldsymbol{PM} }||\cdot||{\hl \boldsymbol{U_M} }-||{\hl \boldsymbol{\tilde{U}_P}}||}.
    \end{aligned} \label{eq_1}
\end{equation}
Here, \( P \) is a target point in the measurement domain and \( S \) is a two-dimensional region surrounding it, containing \( N \) pixels. In other words, \(S\) is the region used to calculate $\Gamma_1$ and $\Gamma_2$ and therefore, \(N\) is equal to the square of the kernel size. \( M \) is a random point in \( S \) and {\hl \( \boldsymbol{z} \)} represents the unit vector perpendicular to the observational surface. \( \theta_M \) denotes the angle between {\hl \( \boldsymbol{U_M}\)} (the velocity vector of point \( M \)) and  \(\boldsymbol{PM}\) (the vector from point \( P \) to point \( M \)). The symbols \(\wedge\), \( \cdot \), and \( \| \cdot \| \) represent the vector cross product, dot product, and norm, respectively.
\( {\hl \boldsymbol{\tilde{U}_P}} = \frac{1}{S} \int_S \boldsymbol{U} \, \mathrm{d}S \) is a local convection velocity around \( P \). \cite{graftieaux2001combining} reported that \( \lvert \Gamma_{1} \rvert \) reaches values ranging from 0.9 to 1 near the vortex center and \( \lvert \Gamma_{2} \rvert \) is equal to \(2/\pi\) at the vortex boundaries. Then, based on these two parameters, the center and boundaries of each vortex can be decided.  

\subsection{Automated Swirl Detection Algorithm} \label{ASDA}
Automated Swirl Detection Algorithm (ASDA) proposed by \cite{liu2019automated} is an automated vortex identification algorithm based on the $\Gamma$-functions method. ASDA contains two essential steps when performing vortex identifications on a data set from observations or simulations. The first step is to estimate the velocity field using Fourier Local Correlation Tracking (FLCT) \citep[][]{welsch2004ilct,fisher2008subsurface}. \cite{liu2019automated} developed an integrated Python wrapper for the FLCT code, which is available at \url{https://github.com/PyDL/pyflct}. Particularly, the pixel width of the Gaussian filter (sigma) is set to 10, and low-pass spatial filtering (kr) and skip are set to None. \cite{fisher2008subsurface} gave a more detailed explanation of other parameters not mentioned here. The next step is applying the $\Gamma$-functions method \citep{graftieaux2001combining} to the velocity field estimated by FLCT.

\cite{liu2019automated} made some minor adjustments to the $\Gamma_{1}$ and $\Gamma_{2}$ functions. For each pixel \(P\), they defined the two parameters as follows:
\begin{equation}
    \begin{aligned}
        &\Gamma_{1}(P) =\boldsymbol{\hat{z}}\cdot\frac{1}{N}\sum_{S}\frac{\boldsymbol{n}_{PM}\times \boldsymbol{v}_{M}}{|\boldsymbol{v}_{M}|}, \\
        &\Gamma_{2}(P) =\boldsymbol{\hat{z}}\cdot\frac{1}{N}\sum_{S}\frac{\boldsymbol{n}_{PM}\times(\boldsymbol{v}_{M}-\overline{\boldsymbol{v}})}{|\boldsymbol{v}_{M}-\overline{\boldsymbol{v}}|}.
    \end{aligned}\label{eq_2}
\end{equation}

The symbols and vectors in Eq.~(\ref{eq_2}) are similar to the corresponding ones in Eq.~(\ref{eq_1}). More detailed interpretations of these functions can be found in relevant previous studies \citep[e.g.,][]{liu2019co-spatial, liu2019automated, liu2019evidence, xie2025photospheric}.

\color{black}
\begin{figure}
    \centering
    \includegraphics[width=\hsize]{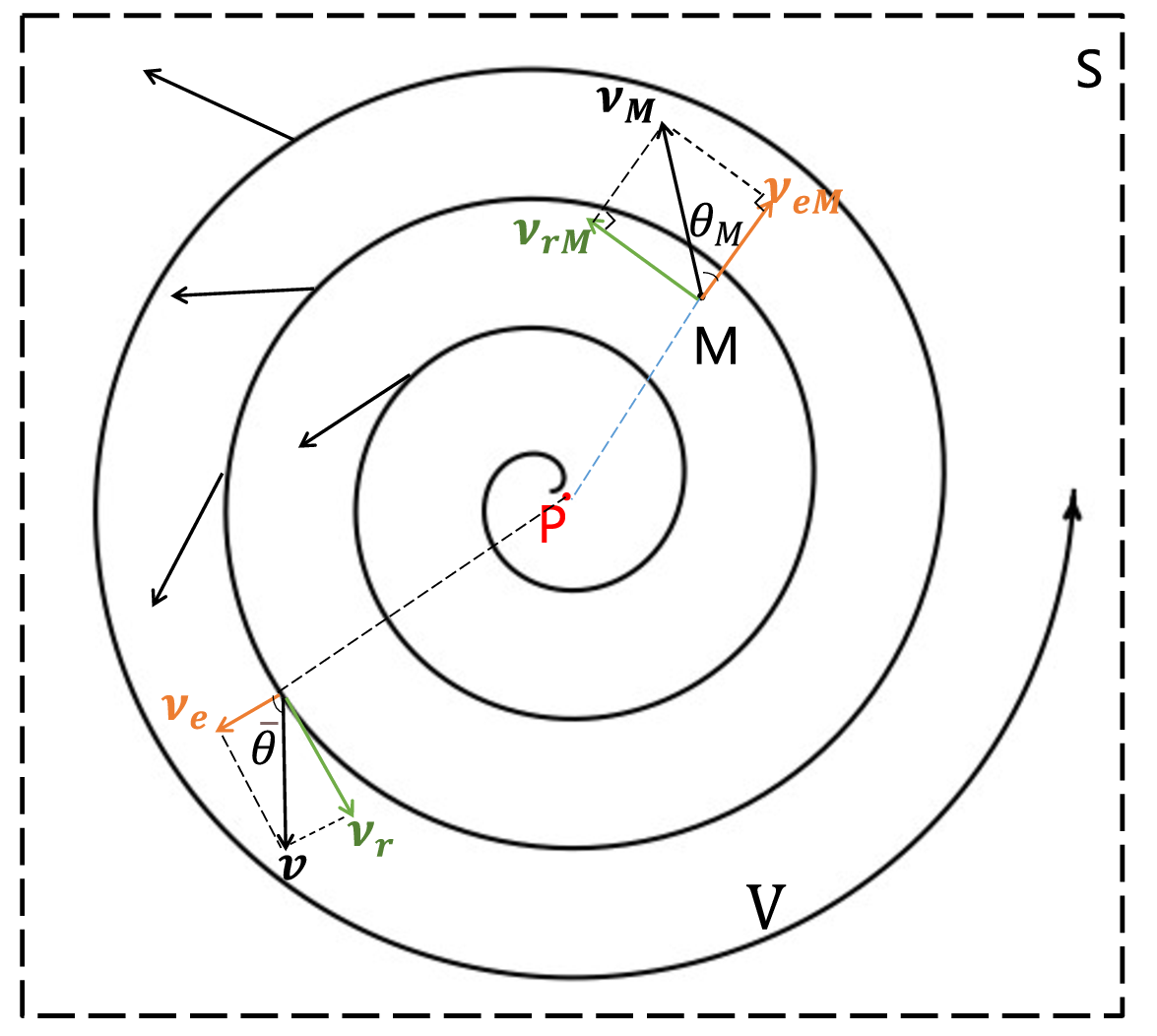}
      \caption{An example vortex \(V\) with its center \(P\). \(M\) is a random point in vortex \(V\), with rotation speed \(v_{rM}\), expansion speed \(v_{eM}\) and speed vector \(\mathbf{v_M}\) at this point. \(S\) is the region used to calculate \(\Gamma_1\) and \(\Gamma_2\) values of the center \(P\). \( v_r \), \( v_e \) and \( \textbf{\textit{v}} \) represent the average rotation speed, expansion speed and speed vector of all points in \(S\), centering on point \(P\). \( \sin \bar \theta \) is defined to the average value of \( \sin (\theta_{M}) \) for all points \(M\) within the region \(S\).}
        \label{vortex-examp}
\end{figure}
   
Figure \ref{vortex-examp} shows an example of a vortex \(V\) with its center \(P\), rotation speed \(v_{rM}\), expansion speed \(v_{eM}\) and speed vector \( \mathbf{v_M} \) at any point \(M\) in region \(S\) used to calculate \(\Gamma_1\) and \(\Gamma_2\) values of its central point \(P\). We note that when $v_e$ is negative, it turns into the contraction speed $v_c$. Meanwhile, \( v_r \), \( v_e \) and \( \textbf{\textit{v}} \) represent the average rotation speed, expansion speed and speed vector of all points in \(S\), centering on point \(P\). \( \sin \bar \theta \) is defined to the average value of \( \sin (\theta_{M}) \) for all points \(M\) within the region \(S\).
\begin{equation}
    \Gamma_{1}(P) = \frac{1}{N}\sum_{S}\sin(\theta_{M}) := \sin \bar \theta ,
\end{equation}
\noindent which means that, \( \bar \theta = \sin ^{-1} \Gamma_1 \). Furthermore, 
\begin{equation}
    \begin{aligned}
        &\frac{v_e}{v_r} = \cot \bar \theta = \cot(\sin ^{-1} \Gamma_1),\\
        &\Gamma_1 = \sin(\cot^{-1}\frac{v_e}{v_r}).
    \end{aligned}
    \label{gamma}
\end{equation}
Therefore, if 0.89 is set as $\Gamma_{1min}$, correspondingly, \( \frac{v_e}{v_r}=\cot(\sin^{-1}0.89) = 0.5\). For a vortex whose \( \frac{v_e}{v_r} > 0.5 \), \( \lvert \Gamma_{1} \rvert \) of its center will be less than 0.89, thus it will not be detected as a vortex by ASDA. For different values of $\Gamma_{1min}$, we can analyze their implications and identify the values of \( \frac{v_e}{v_r} \) of vortices that may be excluded. This analysis will contribute to the subsequent sections of the paper.

\subsection{Validation with Synthetic Data}
To find the optimal parameters of ASDA to detect vortices more accurately, we need to compare the detection results with the exact actual results we know. Therefore, we apply ASDA to synthetic data for comparison, instead of numerical simulation or observational data. In this study, we use Lamb-Oseen vortices \citep{saffman1995vortex} as the synthetic vortices.
\begin{figure}
    \centering
    \includegraphics[width=\hsize]{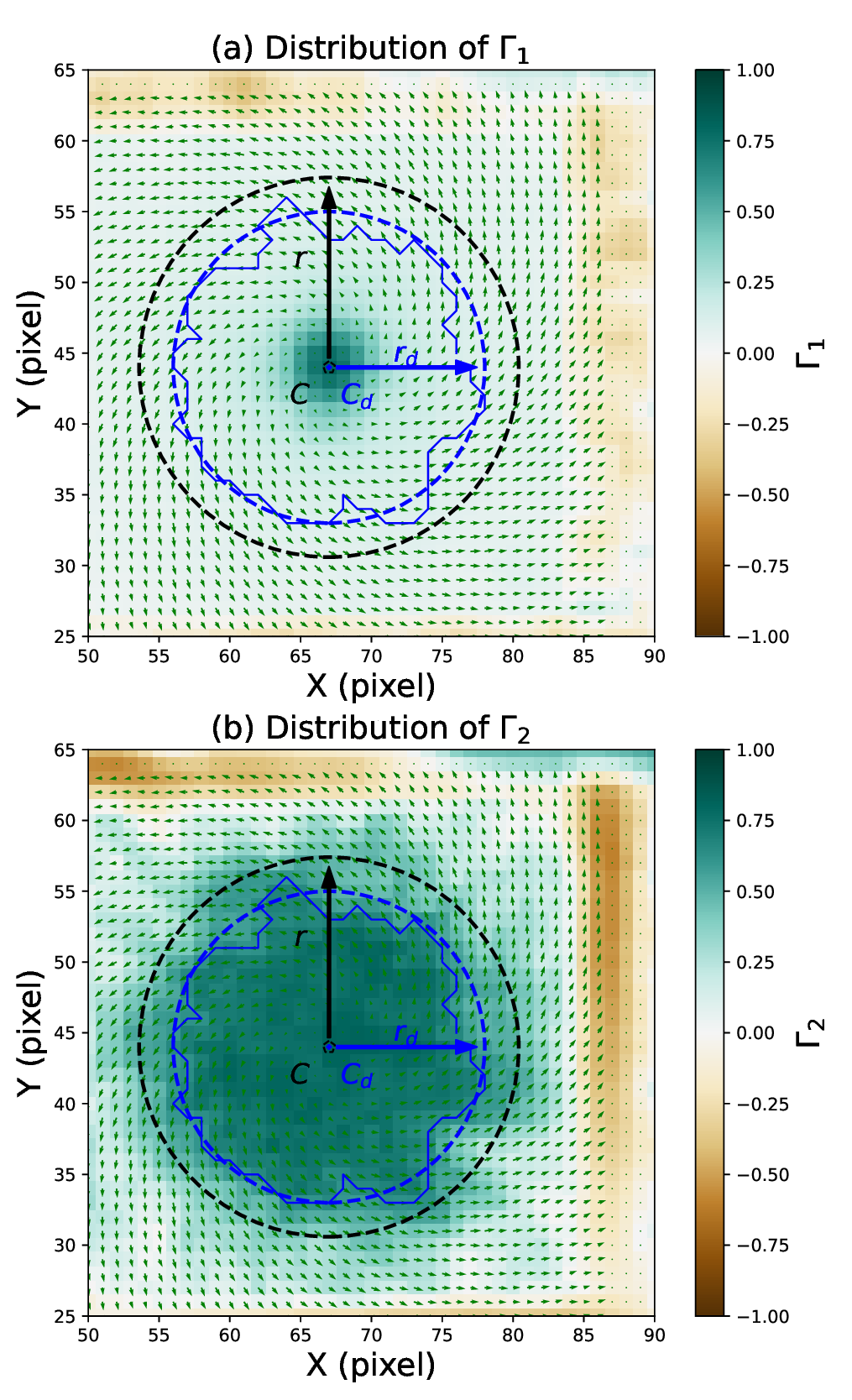}
      \caption{(a) Velocity field of the region (green arrows) and the synthetic vortex edge (black dashed circle) with center \(C\) (black text) and radius \(r\) (black arrow and text). Blue solid curves and blue point ($C_d$) show edge and center of the detected vortex, with effective radius $r_d$ (blue text and arrow). The effective edge (blue dashed circle) is determined by the effective radius $r_d$. The background is the distribution of \(\Gamma_1\). (b) Similar to panel (a) but with the background as the distribution of \(\Gamma_2\).}
      \label{vortex-d-examp}
\end{figure}

Assuming the maximum rotation speed \( v_{max}\) and radius \( r_{max}\) of a Lamb-Oseen vortex, then, a point with a distance \(r\) away from the center has a rotation speed:
\begin{equation}
    v_r=v_{\max}\Bigg(1 + \frac{1}{2\alpha}\Bigg)\frac{r_{\max}}{r}\Bigg[1 - \exp\Bigg(-\alpha\frac{r^2}{r_{\max}^2}\Bigg)\Bigg],
\end{equation}
where, \( \alpha \approx \) 1.256. The expansion/contraction speed \( v_e \) of the vortex can be arbitrarily given. For example, \cite{liu2019automated} chose \( v_e = 0.2v_r\) in their synthetic data. In this way, we can correspond a Lamb-Oseen vortex with its \( \Gamma_1\) value of the center intuitively. Figure~\ref{vortex-d-examp} shows an example Lamb-Oseen vortex. Black dashed curves in panels (a) and (b) are the boundaries and \( r \) is its radius. Blue curves are the boundaries of the vortex detected by ASDA and blue dashed lines show the edges decided by the effective radius \( r_d \) \citep{liu2019automated}. Meanwhile, the center of the Lamb-Oseen vortex \( C \) with coordinate \( (x, y) \) and the detected center \( C_d \) with coordinate (\(x_d\), \(y_d\)) are located at the same pixel in panels (a) and (b) of Figure~\ref{vortex-d-examp}. Following \cite{liu2019automated}, the location accuracy ($A_l$), radius accuracy ($A_r$) and rotation speed accuracy ($A_s$) of the detected vortex are defined as following,
\begin{equation}
    \begin{aligned}
        &A_{l} = (1 - \frac{|(x,y)-(x_{d},y_{d})|}{r}) \times 100\%, \\
        &A_{r} = (1 - \frac{|r_{d}-r|}{r}) \times 100\%, \\
        &A_{s} = (1 - \frac{|v_{rd}-v_{r}|}{|v_{r}|}) \times 100\%.
\end{aligned}
\end{equation}
Here \( v_r \) and \( v_{rd} \) represent the real and calculated rotation speeds of the Lamb-Oseen vortex.

\section{Experiments and Results} \label{experi and res}

\subsection{Optimal \texorpdfstring{$\Gamma_{1min}$}{}} \label{gammamin}

As mentioned in the introduction, there are discrepancies and uncertainties when choosing the value for \( \Gamma_{1min}\) to determine whether a detected feature is a vortex or not. Firstly, we refer to the thoughts of testing ASDA (with a kernel size of 7) using a series of synthetic data in \cite{liu2019automated}. 1000 Lamb-Oseen vortices are generated whose radii \(r_{max}\) and rotation speeds \( v_{max}\) obey the following Gaussian distribution:
\begin{equation}
    f(x)=\frac{1}{\sqrt{2\pi} \sigma}\exp(-\frac{(x-\mu)^2}{2\sigma^2}).
\end{equation}
\noindent Here, \(f(x)\) is the probability density of the variable \(x\), \(\mu\) is its expected value, and \(\sigma\) is the standard deviation. Based on the statistical results of detected photospheric vortices in \cite{liu2019automated}, we set the expected radius of the vortices to \(\mu_r\) = 7.2 pixels with a standard deviation \(\sigma_r\) = 1.5 pixels. Similarly, the expected rotation speed is defined as \(\mu_v\) = 0.17 pixels per frame with a standard deviation of \(\sigma_v\) = 0.07 pixels per frame. For expansion speed \(v_e\) of each vortex, we set \(v_e = \kappa \cdot v_r\). Here, \(\kappa\) is a parameter that also obeys the Gaussian distribution and we set the expected value to \(\mu_{\kappa} = 0.9\) with a standard deviation \(\sigma_{\kappa} = 0.2\). Thus, according to the 3\(\sigma\) rule for Gaussian distributions, approximately 99.7\% of data points fall within \(\pm 3\sigma\) of the mean, indicating that nearly all values of \( \kappa \) are concentrated between 0.3 and 1.5. The range is wide enough to represent most vortices. The generated vortices are then randomly divided into two equal groups: one rotating counterclockwise (positive rotation) and the other clockwise (negative rotation). A background noise map of 5000 \(\times\) 5000 pixel$^2$ is generated, with each pixel assigned a velocity with a random direction and a random magnitude between 0\% and 20\% of \(\mu_v\). Next, the 1000 vortices are randomly placed within this background noise map, ensuring no overlap among them. This process results in a synthetic velocity map (named SD1, synthetic data 1) that closely resembles observational data. Furthermore, we apply ASDA to SD1 using different values of \(\Gamma_{1min}\) from 0.45 to 0.89 to detect vortices. This process is repeated 100 times. Detection results of vortices from SD1 with noise levels of 0 and 20\% are shown in Tables \ref{SD1-0} and \ref{SD1-0.2}.

\begin{table}[ht]
\caption{Average detection rate, false detection rate, location accuracy, radius accuracy, and rotation speed accuracy of the detection on all 1000 inserted vortices in SD1, with a velocity noise level of 0.}
\label{SD1-0}
\centering
\resizebox{0.5\textwidth}{!}{
\begin{tabular}{cccccc}
\toprule
\multirow{2}{*}{$\Gamma_{1min}$} & Detection & False Detection & Location & Radius & Rotation Speed \\
 & Rate & Rate & Accuracy & Accuracy & Accuracy \\
  & \% & \% & \% & \% & \% \\
\midrule
0.45 & 90.7 & 0.0 & 100.0 & 83.6 & 92.8 \\
0.50 & 90.7 & 0.0 & 100.0 & 83.6 & 92.8 \\
0.55 & 90.7 & 0.0 & 100.0 & 83.6 & 92.8 \\
0.60 & 90.7 & 0.0 & 100.0 & 83.6 & 92.8 \\
0.65 & 87.1 & 0.0 & 100.0 & 85.3 & 94.3 \\
0.70 & 65.1 & 0.0 & 100.0 & 90.2 & 96.6 \\
0.75 & 37.9 & 0.0 & 100.0 & 93.4 & 97.7 \\
0.80 & 16.6 & 0.0 & 100.0 & 95.8 & 98.3 \\
0.85 & 5.1 & 0.0 & 100.0 & 97.1 & 98.8 \\
0.89 & 1.3 & 0.0 & 100.0 & 97.4 & 99.1 \\
\bottomrule
\end{tabular}
}
\end{table}
\begin{table}[ht]
\caption{Similar to Table \ref{SD1-0} but with a noise level of 20\%.}
\label{SD1-0.2}
\centering
\resizebox{0.5\textwidth}{!}{
\begin{tabular}{cccccc}
\toprule
\multirow{2}{*}{$\Gamma_{1min}$} & Detection & False Detection & Location & Radius & Rotation Speed \\
 & Rate & Rate & Accuracy & Accuracy & Accuracy \\
  & \% & \% & \% & \% & \% \\
\midrule
0.45 & 92.7 & 0.0 & 98.6 & 72.7 & 86.9 \\
0.50 & 91.2 & 0.0 & 99.1 & 73.5 & 87.3 \\
0.55 & 90.3 & 0.0 & 99.3 & 74.1 & 87.6 \\
0.60 & 88.0 & 0.0 & 99.6 & 75.3 & 88.5 \\
0.65 & 80.8 & 0.0 & 99.8 & 78.5 & 90.9 \\
0.70 & 57.6 & 0.0 & 99.9 & 84.0 & 94.4 \\
0.75 & 31.6 & 0.0 & 99.9 & 88.1 & 96.2 \\
0.80 & 12.8 & 0.0 & 100.0 & 91.5 & 97.3 \\
0.85 & 3.4 & 0.0 & 100.0 & 94.3 & 97.9 \\
0.89 & 0.8 & 0.0 & 100.0 & 96.1 & 98.3 \\
\bottomrule
\end{tabular}
}
\end{table}
Tables \ref{SD1-0} and \ref{SD1-0.2} list the detection rate, false detection rate, location accuracy, radius accuracy and rotation speed accuracy of all detected vortices at velocity noise levels of 0 and 20\% from SD1. There is little difference in results with \( \Gamma_{1min} \) from 0.45 to 0.60: the detection rates and accuracies for the location, radius, and rotation speed all keep high levels. However, both detection rates with the two different noise levels slightly decrease (3.4\% in Table \ref{SD1-0} and 8\% in Table \ref{SD1-0.2}) when \( \Gamma_{1min} \) increases from 0.60 to 0.65, and both of the detection rates drop very quickly when \( \Gamma_{1min} \) is more than 0.65. Note that the false detection rate is consistently zero with increasing \( \Gamma_{1min}  \) and the location accuracy, radius accuracy, and rotation speed accuracy all keep high levels. This finding further proves the result in \cite{liu2019automated} that ASDA is unlikely to detect a vortex at a location where there is none. 

It is mentioned that \( \kappa\) obeys the Gaussian distribution \(N(0.9, 0.2^2)\) for all vortices in SD1, and nearly all values of \( \kappa \) are between 0.3 and 1.5. This means that, theoretically, \(\lvert \Gamma_{1} \rvert\) values of vortices centers should be almost distributed from 0.55 (\(\sin(\cot^{-1}1.5)\)) to 0.96 (\(\sin(\cot^{-1}0.3)\)). Therefore, about 99.7\% vortices could be detected by ASDA with \(\Gamma_{1min}\) = 0.55, and for any \(\Gamma_{1min}\) \(\leq\) 0.55, the detection rates should be very similar. This is consistent with the observation from Tables~\ref{SD1-0} and~\ref{SD1-0.2}. But it is also worth noting that, the detection rates only drop significantly when \(\Gamma_{1min}\) is above 0.65, which may indicate that the optimal value for \(\Gamma_{1min}\) should be around 0.60 to 0.65. Moreover, to avoid occasionality, we conducted two more experiments by setting \(\mu_r\) = 14.4 pixels with \(\sigma_r\) = 2.4 pixels, and \(\mu_r\) = 3.6 with \(\sigma_r\) = 0.8 pixels, respectively. Results for the above two experiments are similar to the result of SD1, suggesting that ASDA performs well in identifying vortices with different radii. 

After considering the radii of vortices, we change the values of \(\kappa\) and take new experiments to explore how the ratio between the expansion and rotation speeds affects the results. A new synthetic data SD2 similar to SD1 also containing 1000 Lamb-Oseen vortices is generated. The expected radius and the standard deviation are also set to \(\mu_r\) = 7.2 pixels and \(\sigma_r\) = 1.5 pixels, respectively, and the expected rotation speed is also defined as \(\mu_r\) = 0.17 pixels with a standard deviation of \(\sigma_v\) = 0.07 pixels. However, for \(v_e = \kappa \cdot v_r\), \(\kappa\) is set to obey the Gaussian distribution \(N(0.5,0.1^2)\) for all 1000 vortices in SD2 (comparing to $\kappa$ obeying \(N(0.9, 0.2^2)\) in SD1). Similarly, there are approximately 99.7\% of values of \(\kappa\) located between 0.2 and 0.8, which indicates that values of \(\lvert \Gamma_{1} \rvert\) at vortices centers are mostly concentrated between 0.78 (\(\sin(\cot^{-1} 0.8)\)) and 0.98 (\(\sin(\cot^{-1} 0.2)\)). There is little difference between the detection rates with \(\Gamma_{1min} \leq 0.78\), as expected, however, an obvious fall occurs when \(\Gamma_{1min}\) increases from 0.75 to 0.80. To further validate these observations, another experiment is conducted with a larger \(\kappa\), obeying the Gaussian distribution \(N(1.2, 0.2^2)\), with all other conditions being kept the same as of SD2, generating the synthetic data SD3. The detection results for SD3 with different \(\Gamma_{1min}\) under noise level of 20\% are shown in Table \ref{SD1_1.2_0.2}.

\begin{table}[ht]
\caption{Similar to Table \ref{SD1-0} but for the detection results of SD3 with a noise level of 20\% }
\label{SD1_1.2_0.2}
\centering
\resizebox{0.5\textwidth}{!}{
\begin{tabular}{cccccc}
\toprule
\multirow{2}{*}{$\Gamma_{1min}$} & Detection & False Detection & Location & Radius & Rotation Speed \\
 & Rate & Rate & Accuracy & Accuracy & Accuracy \\
  & \% & \% & \% & \% & \% \\
\midrule
0.45 & 50.2 & 0.0 & 95.9 & 52.6 & 73.0 \\
0.50 & 47.6 & 0.0 & 97.1 & 54.4 & 73.8 \\
0.55 & 45.9 & 0.0 & 97.7 & 55.6 & 74.6 \\
0.60 & 41.4 & 0.0 & 99.0 & 59.2 & 76.8 \\
0.65 & 29.2 & 0.0 & 99.9 & 68.3 & 84.4 \\
0.70 & 10.3 & 0.0 & 100.0 & 80.3 & 93.3 \\
0.75 & 2.0 & 0.0 & 100.0 & 87.0 & 96.1 \\
0.80 & 0.5 & 0.0 & 100.0 & 91.9 & 97.5 \\
0.85 & 0.1 & 0.0 & 100.0 & 96.4 & 99.4 \\
0.89 & 0.0 &  &  &  &  \\
\bottomrule
\end{tabular}
}
\end{table}
First of all, there is some commonality between the above detection results under different conditions, as seen in Table \ref{SD1-0}, Table \ref{SD1-0.2}, and Table \ref{SD1_1.2_0.2}.  The detection rates all encounter the first ``quick'' decrease from \(\Gamma_{1min}=0.60 \) to \(\Gamma_{1min}=0.65\). The detection rates decrease by 3.4\% in Table \ref{SD1-0}, 8\% in Table \ref{SD1-0.2}, and 12.2\% in Table \ref{SD1_1.2_0.2}. According to Eq.~(\ref{gamma}), values of \(\lvert \Gamma_{1} \rvert\) at vortices centers in SD3 are mostly concentrated between 0.49 (\(\sin(\cot^{-1} 1.8)\)) and 0.86 (\(\sin(\cot^{-1} 0.6)\)). However, the detection rate is only 50.2\% with \(\Gamma_{1min} = 0.45\), significantly less than 99.7\%. It indicates that some vortices are detected as candidate vortices by criterion \(\Gamma_{1min}\) but rejected by other criteria. Based on the methodology of ASDA mentioned in Sect. \ref{ASDA}, we speculate that the criterion on \(\Gamma_2\) might have also omitted some candidate vortices. To verify this, we sample a small region (200 \(\times\) 200 pixel$^2$) of SD3 to compare the location distribution of vortices with the value distribution of \(\Gamma_{1}\) and \(\Gamma_{2}\).

\begin{figure*}
   \centering
   \includegraphics[width=1.0\textwidth]{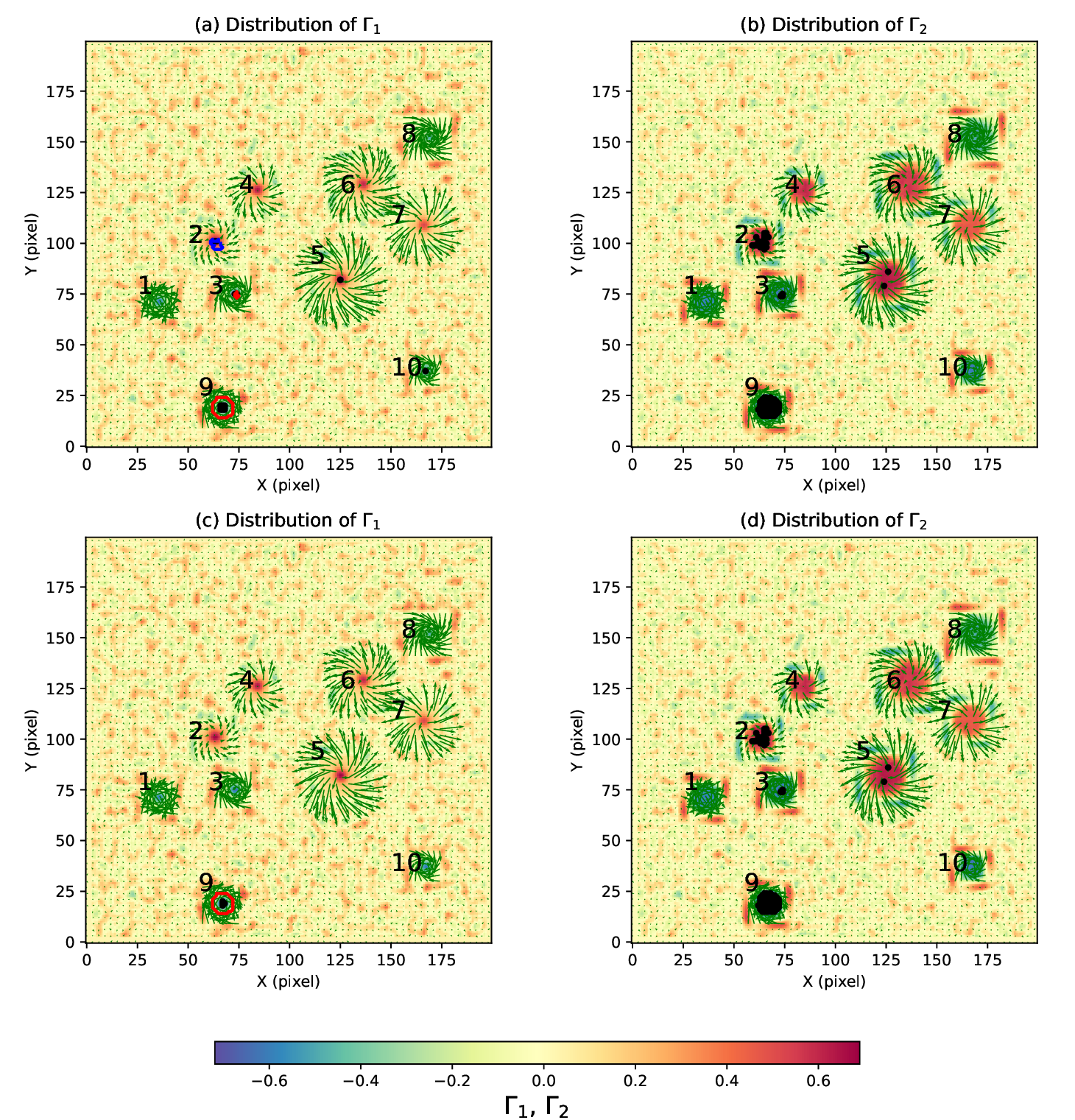}
   \caption{(a) and (b) are the distributions of \(\Gamma_1\) and \(\Gamma_2\) (backgrounds) in a 200 \(\times\) 200 pixel$^2$ region of SD2. The green arrows represent the velocity field, and the numbers label the 10 synthetic lamb-Oseen vortices. The black dots in (a) are the points where \(\lvert \Gamma_1 \rvert\) are more than 0.60, and black dots in (b) correspond to points where \(\lvert \Gamma_2 \rvert\) are more than 2/$\pi$. (c) and (d) are similar to (a) and (b), respectively. But the black dots in (c) and (d) show the points where \(\lvert \Gamma_1 \rvert\) are more than 0.65 and \(\lvert \Gamma_2 \rvert\) are more than 2/$\pi$, correspondingly. The blue and red curves in these four panels represent the boundaries of vortices rotating counterclockwise and clockwise, respectively.}
    \label{Gamma2}%
\end{figure*}

There are a total of 10 synthetic vortices in the 200 \(\times\) 200 pixel$^2$ region, which are marked using black numbers in the four panels of Figure~\ref{Gamma2}. The blue and red curves represent the boundaries of the detected vortices rotating counterclockwise and clockwise, respectively. Panels (a) and (b) are the detection results using ASDA with \(\Gamma_{1min}\) = 0.60, panels (c) and (d) represent the results of detection results with \(\Gamma_{1min}\) = 0.65. Backgrounds in Figure~\ref{Gamma2} are the distributions of \(\Gamma_1\) and \(\Gamma_2\). Green arrows represent the velocity fields and black dots in panels (a) and (c) show where values of \(\lvert\Gamma_1\rvert \geq 0.60\) and values of \(\lvert\Gamma_1\rvert \geq 0.65\), respectively. Meanwhile, black dots in panels (b) and (d) represent the points with values of \(\lvert\Gamma_2\rvert \geq 2/\pi\). One can see from Figure~\ref{Gamma2}(a) and (c) that vortices Nr. 1, 4, 5, 6, 7, 8 and 10 are denied as vortices with \(\Gamma_{1min}\) = 0.60 and 0.65, among which Nr. 1, 4, 6, 7 and 8 are also denied by \(\Gamma_2\) (Fig.~\ref{Gamma2}b and d). However, Nr. 5 and Nr. 10 pass the condition \(\lvert\Gamma_1\rvert \geq 0.60\) though it is denied by \(\lvert\Gamma_1\rvert \geq 0.65\) and \(\lvert\Gamma_2\rvert \geq 2/\pi\). This indicates that if we set \(\Gamma_{1min}\) to 0.60 or less, some vortices will be accepted by the \(\Gamma_1\) condition but refused by the \(\Gamma_2\) condition. Although it makes little influence to the detection results but may cause a massive waste of computing resources for a huge dataset. 

Secondly, for vortices Nr. 2 and 3, they are detected as a positive vortex and a negative vortex, respectively, with \(\Gamma_{1min} = 0.60\). However, they are omitted by ASDA with \(\Gamma_{1min} = 0.65\), because \(\lvert \Gamma_1 \rvert\) values of points in vortices Nr. 2 and 3 are all less than 0.65 and no points are identified as their centers. Note that their boundaries are both detected successfully (see black dots of Nr. 2 and 3 in Fig.~\ref{Gamma2}b and d) according to the $\Gamma_2$ criterion \citep[i.e., \( \lvert \Gamma_{2} \rvert = 2/\pi\) at the vortex boundaries][]{graftieaux2001combining}. Then, it is concluded that vortices Nr. 2 and 3 are both real vortices and when we set \(\Gamma_{1min} = 0.65\) or larger, they would not be detected as vortices. In other words, if \(\Gamma_{1min}\) is too large (0.65 or larger), some real vortices would not be detected by ASDA and therefore we will underestimate the number of vortices.

In conclusion, we can first determine whether a rotational structure is a real vortex based on the \(\Gamma_2\) criterion. If its boundaries where \(\lvert \Gamma_2 \rvert = 2/\pi\) are identified, an optimal \(\Gamma_{1min}\) should be used to determine whether the candidate is a vortex or not. The above experiments with different distributions of vortices and different levels of noise suggest that the optimal \(\Gamma_{1min}\) should be between 0.60 and 0.65. To find the exact value of the optimal \(\Gamma_{1min}\), we apply ASDA with \(\Gamma_{1min} = 0.45\) (enough small) to SD1 (\(v_e = N(0.9, 0.2^2)\cdot v_r\)) and SD3 (\(v_e = N(1.2, 0.2^2)\cdot v_r\)) with both noise levels of 0. This makes the value range of \(v_e / v_r\) (0.3 \(\sim\) 1.8) wide enough to cover most distributions of \(v_e / v_r\). Statistically, the minimum values of \( \lvert \Gamma_{1} \rvert\) of vortices centers detected in SD1 and SD3 are 0.638 and 0.637, respectively. The values are both between 0.60 and 0.65, which can explain the quick decreases from \(\Gamma_{1min}=0.55\) to \(\Gamma_{1min}=0.60\) in Tables \ref{SD1-0}, \ref{SD1-0.2} and \ref{SD1_1.2_0.2}. Therefore, we suggest the optimal value of \(\Gamma_{1min}\) is 0.63, and we can detect almost all vortices applying ASDA with \(\Gamma_{1min} = 0.63\). This optimal value of \(\Gamma_{1min}\) will be further tested and validated with numerical simulation and observational data in the rest of this section.

\subsection{Influence of the kernel size} \label{ks}
After considering the optimal value of \(\Gamma_{1min}\) is 0.63, in this subsection, let us now explore the influence of the kernel size ($ks$) in calculating \(\Gamma_1\) and \(\Gamma_2\). \cite{liu2019automated} specified $N$ ($ks^2$) as 49, which means that they selected 7 as the kernel size and calculated \(\Gamma_1\) and \(\Gamma_2\) in a region of 7 \(\times\) 7 pixel$^2$ surrounding each target point. The above specification will weaken vortices with radii less than 4 pixels, which indicates that ASDA may fail to identify some small-scale vortices (radii less than 4 pixels) and would underestimate the number of vortices.

\cite{yuan2023advanced} proposed the Advanced \(\Gamma\) Method (AGM) to identify vortices and used an adaptive version to optimize AGM for vortex identification. The adaptive version is based on a sequence of different kernel sizes, such as 3, 5, 7, 9, 11 and so on. \cite{yuan2023advanced} noted that there are different values of \(\Gamma_1\) and \(\Gamma_2\) for the same point using different kernel sizes to calculate. For example, if \(\lvert \Gamma_1 \rvert\) of a point is calculated less than \(\Gamma_{1min}\) using $ks$ = 7 but more than \(\Gamma_{1min}\) using $ks$ = 9, then, this point may be a potential vortex center and should not be omitted immediately. Therefore, \cite{yuan2023advanced} calculate values of \(\Gamma_1\) with several kernel sizes (3, 5, 7, 9, and 11) and use the maximum \(\lvert \Gamma_1 \rvert\) at each pixel under different kernel sizes. \cite{yuan2023advanced} also suggested that varying the kernel size for each vortex provides better identification and leads to more accurate statistical results of the vortex parameters. However, experiments carried out by \cite{yuan2023advanced} were based on an arbitrary \(\Gamma_{1min}\) = 0.75, and how the different kernel sizes influence the detection results with the optimal \(\Gamma_{1min}\) = 0.63 needs further examination. 

In this subsection, we first calculate \(\Gamma_1\) and \(\Gamma_2\) with kernel sizes = 3, 5, 7, 9, and 11. At each pixel, only the maximal values of \(\lvert \Gamma_1 \rvert\) and \(\lvert \Gamma_2 \rvert\) with different kernel sizes are combined. Swirl detection based on ASDA is then applied to these combined \(\Gamma_1\) and \(\Gamma_2\) values from different kernel sizes. We name this version VGCM-o (variable $\Gamma$ calculating method-origin) and test it with synthetic data with \(\Gamma_{1min} =  0.63\). According to Eq. (\ref{gamma}), \(\Gamma_{1min} = 0.63\) corresponds to \(\kappa = v_e / v_r = 1.23\), meaning that vortices whose \(\kappa > 1.23\) will be not detected by ASDA. Therefore, to avoid the potential influence of \(\Gamma_{1min}\), SD2 is the most suitable dataset to test VGCM-o with \(\Gamma_{1min} =  0.63\), because \(\kappa\) values of vortices in SD2 are mostly located between 0.2 and 0.8. The detection results of SD2 with noise levels ranging from 0 to 20\% are shown in Table \ref{SD1-1.2-0.2-VGM-o}.

\begin{table}[ht]
\caption{Average detection rate, false detection rate, location accuracy, radius accuracy, and rotation speed accuracy of the detection on all 1000 inserted vortices in SD2 applying VGCM-o with \(\Gamma_{1min} = 0.63\), with a velocity  noise level ranging from 0 to 20\%.}
\label{SD1-1.2-0.2-VGM-o}
\centering
\resizebox{0.5\textwidth}{!}{
\begin{tabular}{cccccc}
\toprule
Noise & Detection & False Detection & Location & Radius & Rotation Speed \\
Level & Rate & Rate & Accuracy & Accuracy & Accuracy \\
  & \% & \% & \% & \% & \% \\
\midrule
0 & 100.0 & 0.0 & 100.0 & 97.0 & 99.7 \\
5\% & 113.9 & 12.6 & -31.3 & 86.3 & 86.6 \\
10\% & 116.2 & 14.5 & -86.9 & 84.8 & 84.3 \\
15\% & 113.8 & 13.1 & -47.9 & 85.5 & 85.2 \\
20\% & 114.1 & 13.8 & -50.5 & 84.4 & 83.0 \\
\bottomrule
\end{tabular}
}
\end{table}
Table~\ref{SD1-1.2-0.2-VGM-o} shows that the detection maintains high accuracy for the location, radius, and rotation speed of vortices at a noise level of 0. However, when there is noise, even with a noise level of only 5\%, the detection rate is above 100\% and there are false detections. The location accuracy turns negative, which indicates that the detected vortex center is outside the synthetic vortex. The radius accuracy and rotation speed accuracy still keep high levels. To confirm these findings, we also change the radii of vortices in SD2 to larger and smaller values and still obtain similar detection results. This suggests that VGCM-o performs well in identifying vortices when there is no noise, but detects some false vortices and yields unreliable vortex centers when noises exist in the dataset, which is very common when the dataset is obtained from observations. 

To study the reasons for the poor behavior of VGCM-o with noisy data and search for a better method to calculate \(\Gamma_1\) and \(\Gamma_2\), we recalculate the values of \(\Gamma_1\) and \(\Gamma_2\) of SD2 using different single kernel sizes (ranging from 3 to 15) and apply detection parts of ASDA to these \(\Gamma_1\) and \(\Gamma_2\) values, also with \(\Gamma_{1min} = 0.63\).

The detection results are shown in Table \ref{SD2-0.9-0.2-diff-ks}. We find that the detection rate is highest when $ks$ = 9 and the detection rate turns lower if $ks$ increases (9, 11, 13, and 15) or decreases (5 and 7). Moreover, only when $ks$ = 3, the detection rate is more than 100\%, and false vortices are detected, with very poor location, radius, and rotation speed accuracies. It is clearly shown that the detection results with VGCM-o and $ks$ = 3 are similar and therefore, the bad detection results with VGCM-o should result from $ks$ = 3. To verify this, we revise VGCM-o by removing $ks$=3 when calculating \(\Gamma_1\) and \(\Gamma_2\). 

In Table~\ref{SD2-0.9-0.2-diff-ks}, VCGM-1 uses kernel sizes of 5, 7 and 9, VCGM uses kernel sizes of 5, 7, 9 and 11, and VGCM-2 uses kernel sizes of 5, 7, 9, 11 and 13. It is shown that for these three combinations of different kernel sizes, the detection rates are all at high and reasonable levels (< 100\%), with high location, radius and rotation speed accuracies. Not a single false vortex has been detected, indicating that the detection results are better after abandoning $ks$ = 3. Moreover, the results of VGCM, VGCM-1, and VGCM-2 are all better than the results of using any single kernel size, which proves the variable \(\Gamma\)-functions method does contribute to the more accurate detection of vortices. Considering that the detection rates of VGCM and VGCM-2 are the same and both 0.3\% higher than the detection rate of VGCM-1, VGCM (with kernel sizes of 5, 7, 9 and 11, and costing less computation power compared to VGCM-2) is found to be the best for SD2 whose vortices radii follow the Gaussian distribution of  N(7.2, 1.6$^2$). 

Next, let us vary the radii of vortices in SD2 to smaller and larger values following N(3.6, 0.8$^2$) and N(14.4, 2.4$^2$), respectively, and repeat the above experiments. Both results support our findings that the bad detection results with VGCM-o result from $ks$ = 3. When only using a single kernel size, $ks$ = 5 is the best choice for smaller vortices but $ks$ = 19 is the best for larger vortices. This suggests that the best single kernel size is always close to the average radius of vortices in the dataset. Although we also find that a Variable \(\Gamma\) Calculating Method containing this best single kernel size performs better, the improvement compared to VGCM (kernel sizes = 5, 7, 9 and 11) is very little ($\sim$1\%) but costs significantly more computing resources. Moreover, because it is impossible to know the average radius of vortices in observational data, we can not adjust the Variable \(\Gamma\) Calculating Method by selecting specific kernel sizes. Therefore, in practice, VGCM (with kernel sizes of 5, 7, 9 and 11) is most suitable to detect vortices accurately and meanwhile avoid false detections. 

\begin{table}[ht]
\caption{Average detection rate, false detection rate, location accuracy, radius accuracy, and rotation speed accuracy of the detection on all 1000 inserted vortices in SD2 applying different kernel sizes and different versions of the Variable \(\Gamma\) Calculating Method (VGCM) with \(\Gamma_{1min} = 0.63\), with a velocity noise level of 20\%.}
\label{SD2-0.9-0.2-diff-ks}
\centering
\resizebox{0.5\textwidth}{!}{
\begin{tabular}{cccccc}
\toprule
Kernel & Detection & False Detection & Location & Radius & Rotation Speed \\
Size & Rate & Rate & Accuracy & Accuracy & Accuracy \\
  & \% & \% & \% & \% & \% \\
\midrule
3 & 105.5 & 11.8 & -26.8 & 70.3 & 77.3 \\
5 & 96.0 & 0.0 & 99.6 & 88.7 & 95.9 \\
7 & 97.2 & 0.0 & 99.8 & 91.3 & 96.7 \\
9 & 98.0 & 0.0 & 99.8 & 90.0 & 95.9 \\
11 & 97.9 & 0.0 & 99.8 & 85.0 & 93.1 \\
13 & 96.4 & 0.0 & 99.6 & 77.7 & 89.0 \\
15 & 93.4 & 0.0 & 99.3 & 70.0 & 84.3 \\
VGCM-o & 114.1 & 13.8 & -50.5 & 84.4 & 83.0 \\
VGCM-1 & 98.1 & 0.0 & 99.8 & 94.3 & 97.7 \\
VGCM & 98.4 & 0.0 & 99.9 & 95.0 & 97.8 \\
VGCM-2 & 98.4 & 0.0 & 99.9 & 95.3 & 98.0 \\
\bottomrule
\end{tabular}
}
\end{table}
\subsection{Validation with numerical simulation data} \label{simulation}
\subsubsection{Photosphere} \label{ptsc simulation}
\begin{figure*}
   \centering
   \includegraphics[width=1.0\textwidth]{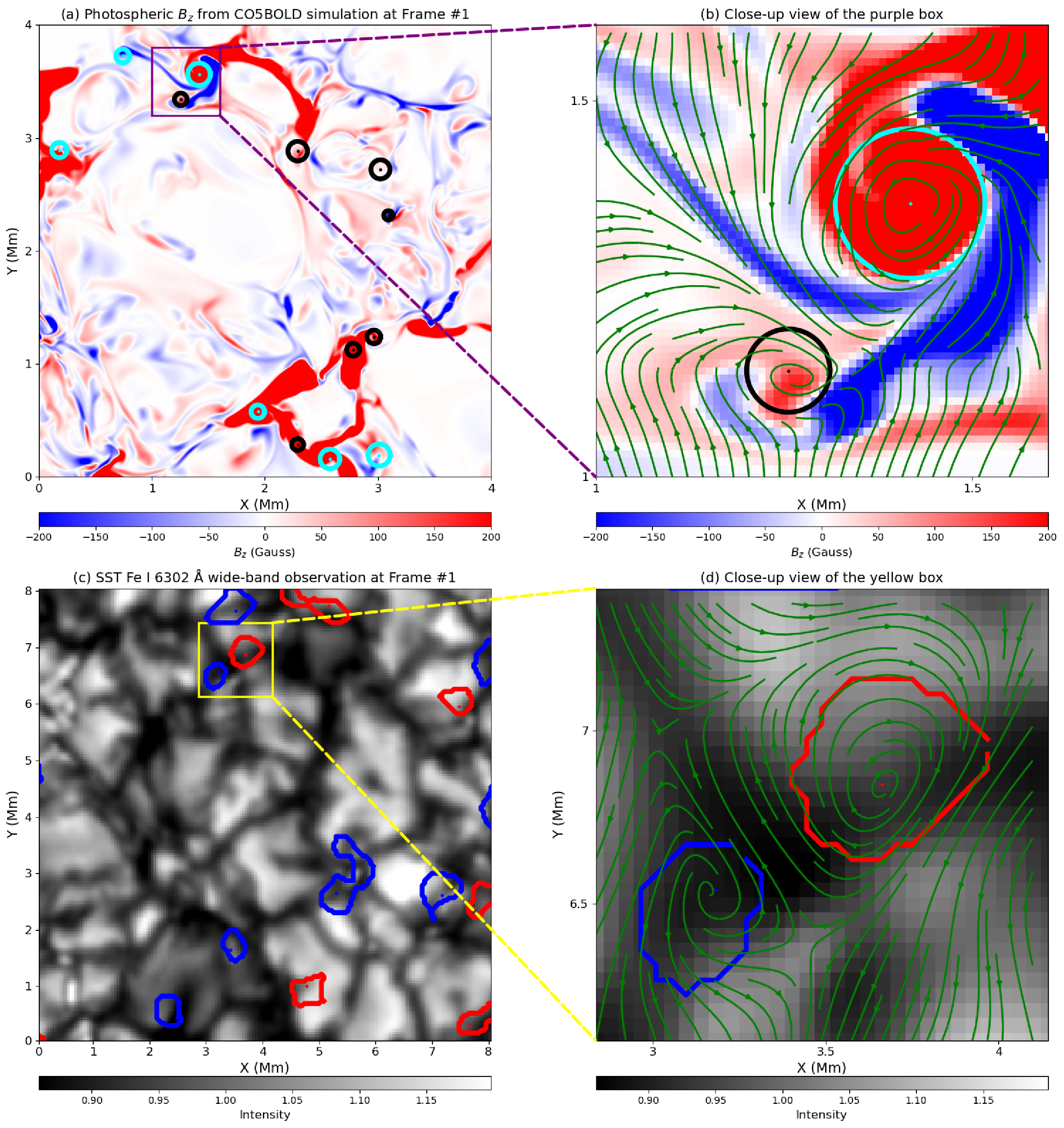}
   \caption{{\hl (a) and (c) are both the first frame of CO5BOLD simulation data and SST observational data, showing the $B_z$ from CO5BOLD simulation and the photospheric intensity from SST observations, respectively. The cyan (blue) and black (red) curves in panel (a) (panel (c)) denote the boundaries of counter-clockwise and clockwise vortices detected by SWIRL (ASDA). (b) and (d) show the close-up views of the purple box in panel (a) and the yellow box in panel (c), respectively. Green arrows in (b) and (d) represent the velocity field.}}
    \label{vortex_exam_b}%
              
\end{figure*}

In Sect. \ref{gammamin} and \ref{ks}, we concluded that VGCM is a suitable approach to calculate \(\Gamma_1\) and \(\Gamma_2\) and $\Gamma_{1min}$ = 0.63 is a more practical choice than 0.89 \citep{liu2019automated} and 0.75 \citep{yuan2023advanced}. We name this version of the improved ASDA, which calculates \(\Gamma_1\) and \(\Gamma_2\) using VGCM and detects vortices with \(\Gamma_{1min}\) = 0.63, as the Optimized ASDA. We note that the above results are obtained by experiments with various synthetic data, and whether Optimized ASDA can be applied to observational data is to be studied. Therefore, before applying Optimized ASDA to observational data, let us test it with advanced numerical simulation obtained with the CO5BOLD code \citep{freytag2012simulations}. CO5BOLD has been widely used to model stellar atmospheres, such as the Sun, solar-type stars, red giants, and white and brown dwarfs \citep{straus2017third}. Different solvers, such as a hydrodynamic module or a magnetohydrodynamic module, and radiative transfer schemes can be chosen to simulate variable situations, due to a modular construction of the code \citep{straus2017third}. 

\begin{figure*}
   \centering
   \includegraphics[width=1.0\textwidth]{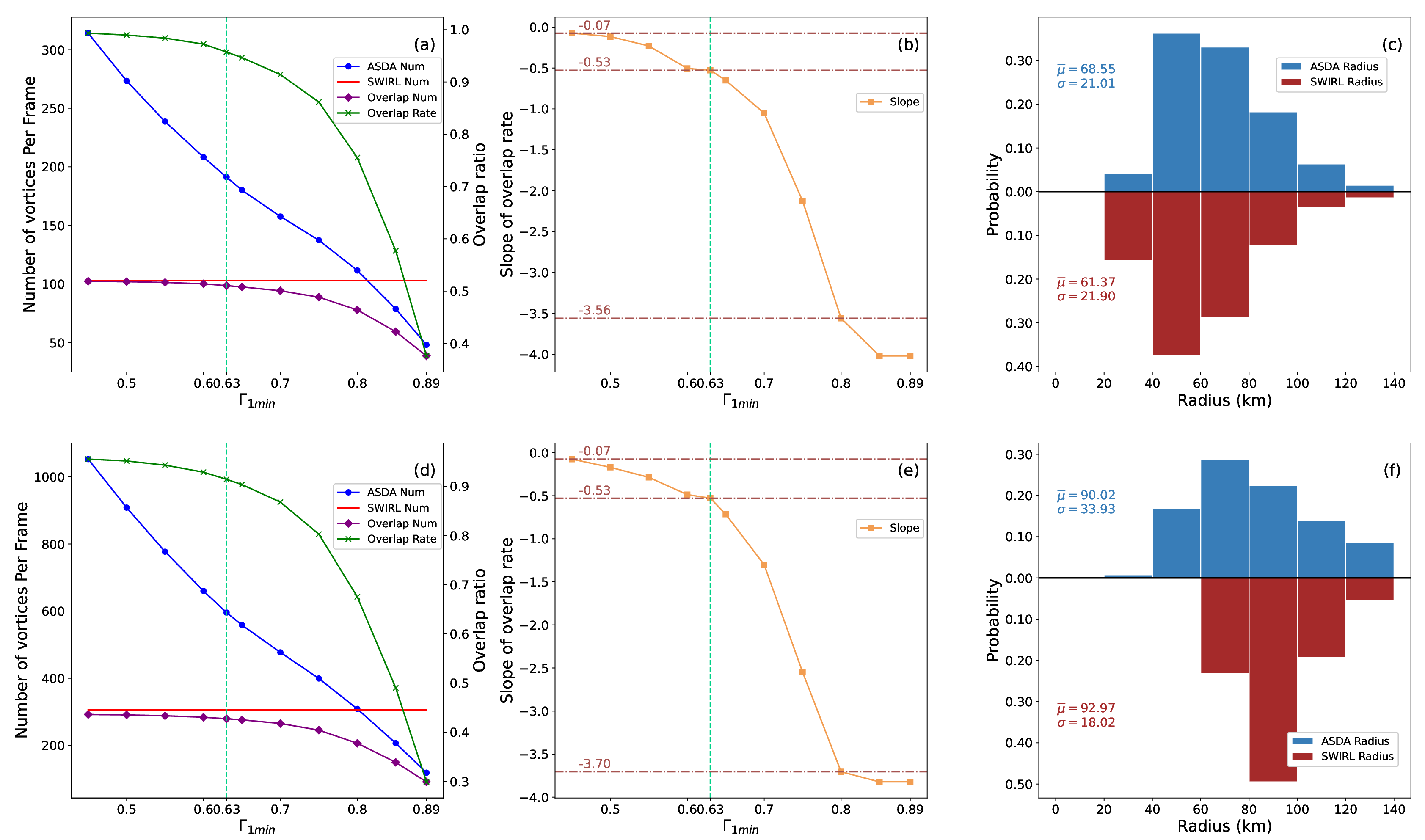}
   \caption{Comparisons between the detection results obtained from the Optimized ASDA and SWIRL with numerical simulations. Panels (a), (b), and (c) are detection results from the photospheric simulation data. The blue curve and red horizontal line in panel (a) show the average numbers of vortices per frame detected by the Optimized ASDA and SWIRL, respectively. The purple curve represents the number of overlapping vortices detected by the Optimized ASDA and SWIRL. The green curve is the corresponding overlapping rate. Panel (b) depicts the slope of each point at the green curve in panel (a). The blue and red histograms in panel (c) denote the distributions of the radius of the overlapping vortices detected by the Optimized ASDA and SWIRL. Panels (d), (e), and (f) are similar to (a), (b), and (c), respectively, but are the results of the chromospheric simulation data.}
    \label{co5bold}%
              
\end{figure*}

In this subsection, we use data from the radiative MHD code for the simulation of the surface layers of the Sun. The simulation was based on a relaxed, purely hydrodynamical model with an initial vertical and homogeneous magnetic field of 50 G. The HLLMHD solver \citep{doi:10.1137/1025002, schaffenberger2005magnetohydrodynamic, schaffenberger2006holistic} was used to ensure the positivity of the gas pressure. The magnetic and plasma boundary conditions were both periodic at the sides, while the magnetic field was enforced to be vertical at the top and bottom of the box. The Cartesian simulation box has a grid spacing of 960 \(\times\) 960 \(\times\) 280 grid cell$^3$ and the cell size is 10 km in each spatial direction, which represents a total size of 9.6 \(\times\) 9.6 \(\times\) 2.8 Mm$^3$. The height of box (labelled as $z$) ranges from -1240 km to 1560 km, with \( z = 0 \) km representing the average optical depth \(\tau_{500} = 1\). Therefore, the simulation domain encompasses layers near the solar surface, including the convection zone, photosphere, and up to the middle chromosphere \citep{cuissa2024innovative}. This simulation started at t = 0 s and run for about 7680 s (i.e. about 2.1 h), with a cadence of 240 s. Discarding the first 1600 s of the simulation (this is typically the time for the initial magnetic field to relax), 26 data cubes were obtained, from t = 1680 s to t = 7680 s. 

\cite{cuissa2022innovative} proposed an innovative and automated method for vortex identification, named SWIRL. This algorithm mainly involves two steps: (1) estimating the vortex center map for each image and, (2) clustering the estimated centers and deciding the vortices. In more detail, for a point with coordinate \( (x, y) \) and velocity \( (v_x, v_y) \), the vorticity \( \mathcal{\omega} \), the velocity gradient tensor \( \mathcal{U} \), the real eigenvector \( \mathbf{u_r} \), the swirling strength \( \lambda \), and the Rortex \( R \) can be computed based on the definitions in \cite{cuissa2022innovative}. Then, we can calculate the radial direction and curvature radius of this point, which can decide its estimated vortex center (EVC). 

Applying this method, one can get the EVC map of each image. Furthermore, based on the EVC maps, the number of EVCs in each grid cell (EVC density) can be counted using the clustering by fast search and finding of density peaks (CFSFDP) algorithm proposed by \cite{rodriguez2014clustering}. Then, cluster EVCs are obtained based on some criteria (see details in \cite{cuissa2022innovative}), and candidate vortices are identified. After noise removal, identified vortices and noisy grid cells are distinguished. Based on this, several properties of each vortex can be obtained, including its center coordinate, effective radius defined by \cite{cuissa2022innovative}, and rotational direction (counter-clockwise or clockwise). More details about the SWIRL method can be found in \cite{cuissa2022innovative}. \cite{cuissa2024innovative} recommended a set of SWIRL algorithm parameters used to detect vortices in the above simulation data cubes. Next, we carry out comparisons between the vortex detection results by Optimized ASDA and SWIRL applied to the photospheric velocity field of the numerical simulation by \cite{cuissa2022innovative}.

{\hl Panels (a) and (b) in Figure~\ref{vortex_exam_b} show an example $B_z$ from the first frame of the CO5BOLD simulations at the photosphere. Cyan and black curves depict the boundaries of counter-clockwise and clockwise vortices identified by SWIRL, with the purple box shown in detail in panel (b).}
Here, when applying Optimized ASDA, we still employ VGCM to calculate \(\Gamma_1\) and \(\Gamma_2\) but vary the value of \(\Gamma_{1min}\) from 0.45 to 0.89 to explore how the vortex detection would be affected with different \(\Gamma_{1min}\) criteria. Detection results from Optimized ASDA and SWIRL are shown in Figure \ref{co5bold}. 

In Figure~\ref{co5bold}(a), the blue curve shows the average number of photospheric vortices per frame detected by Optimized ASDA under different \(\Gamma_{1min}\) values from the 26 photospheric simulation data cubes. The number decreases almost linearly as \(\Gamma_{1min}\) increases. The number of vortices detected by SWIRL is depicted with the horizontal red line. The purple curve presents the number of vortices detected by both Optimized ASDA and SWIRL. It is seen that when \(\Gamma_{1min}\) is less than 0.5, almost all vortices detected by SWIRL are also detected by the Optimized ASDA. However, the number of overlapping decreases slowly with increasing values of \(\Gamma_{1min}\) when \(\Gamma_{1min}\) is less than 0.65, above which, the overlapping number decreases rapidly. To understand the decreasing tendency of the number of overlapping vortices more clearly, we calculate the overlap rates, which is defined by the percentage of overlapping vortices over the total number of vortices detected by SWIRL (green curve in Fig.~\ref{co5bold}a). Figure~\ref{co5bold}(b) shows the slope at each point along the green curve in Figure~\ref{co5bold}(a). The slopes of the overlap rate at \(\Gamma_{1min}\) = 0.45, 0.63 and 0.80 are -0.07, -0.53 and -3.56, respectively. The two bright teal dotted vertical lines in panels (a) and (b) both represent the results of \(\Gamma_{1min} = 0.63\).  The overlap rate decreases much more (panel a) and more quickly (panel b) from \(\Gamma_{1min}\) = 0.63 to 0.80 than from \(\Gamma_{1min}\) = 0.45 to 0.63. This suggests that most vortices detected by SWIRL can also be identified by Optimized ASDA with \(\Gamma_{1min}\) = 0.63 or less. But, when \(\Gamma_1\) becomes larger, Optimized ASDA will miss a significant number of vortices thus underestimating the number of vortices in the data. These results are consistent with what we have obtained from the synthetic data in Sect.~\ref{gammamin} and Sect.~\ref{ks}, further supporting that 0.63 is an optimal choice for \(\Gamma_{1min}\). 

We also note that \cite{liu2019automated} and \cite{cuissa2022innovative} computed the effective radius of vortices using the same method, which is defined as the radius of a circle that has the same area as the vortex: 
\begin{equation}
    R_{eff} = \sqrt{\frac{A_{eff}}{\pi}}.
\end{equation}
Here, \(A_{eff}\) is the effective area of a vortex, decided by the number of grid cells within the vortex and the size of cells. Therefore, we pay attention to the distributions of the radii of those vortices which are detected by both Optimized ASDA and SWIRL. These distributions are shown in Figure \ref{co5bold}(c), with little difference between the distributions of the radii of vortices detected by Optimized ASDA and SWIRL. The expected values of the vortex radius detected by Optimized ASDA and SWIRL are 71.05 km and 61.37 km, with their corresponding standard deviations of 19.88 km and 21.90 km, respectively. The above results suggest that Optimized ASDA with \(\Gamma_{1min}  = 0.63\) can not only detect most vortices identified by SWIRL (and more vortices ignored by SWIRL) but also perform very well in determining the radii of the detected vortices. These again support our previous findings about Optimized ASDA from synthetic data.

\subsubsection{Chromosphere} \label{cmsc simulation}
In this subsection, we explore the performance of Optimized ASDA by applying it to the chromospheric data in the above-mentioned numerical simulation. The utilized chromospheric simulation data cubes are in the same horizontal domain and sampled at the same time as those chosen in Sect.~\ref{ptsc simulation}. However, the height corresponding to the bottom of the chromosphere is at \(z\) = 700 km, higher than the height (\(z\) = 100 km) of photospheric data cubes in Sect.~\ref{ptsc simulation}.

The results are similar to those obtained from the photospheric simulation data cubes, shown in Figure~\ref{co5bold}(d)-(f). It is seen in panel (d) that the blue curve, representing the number of vortices detected by Optimized ASDA, also experiences an almost linear decrease with increasing values of \(\Gamma_{1min}\). The purple and green curves are also very similar to their corresponding curves in panel (a), although more vortices have been detected by both Optimized ASDA and SWIRL from the chromospheric data. Similar to panel (b), panel (e) depicts the slope of each point along the green curve in panel (d), and the slope of overlap rates are -0.07, -0.53 and -3.70 at \(\Gamma_{1min}\) = 0.45, 0.63 and 0.80, respectively. The difference between the slopes from \(\Gamma_{1min}\) = 0.45 to 0.63 is neglectable compared to the variation of the slope with \(\Gamma_{1min}\) from 0.63 to 0.80. This suggests that the above analysis on chromospheric detection results is consistent with results from the photospheric simulation data and also supports the conclusion that 0.63 is an optimal value for \(\Gamma_{1min}\). {\hl Moreover, the radii of overlapping vortices detected by the Optimized ASDA seem to be slightly smaller than the radii detected by SWIRL}, as shown in Figure \ref{co5bold}(f). The expected values of the radii detected by Optimized ASDA and SWIRL are 98.17 km and 92.97 km, with the corresponding standard deviations of 37.78 km and 18.02 km, respectively. 

Comparing vortices detected from the photosphere and chromosphere suggests that there are more vortices in the solar chromosphere than the photosphere in the numerical simulation. This is consistent with the observational fact that more vortices are detected by ASDA from the chromospheric observations than from the photospheric observations \citep{liu2019evidence}. \cite{cuissa2024innovative} suggested that the growth of vortex radii could be explained by the steep decrease in mass density from the photosphere to the chromosphere, which results in the expansion of the plasma ascending into the chromospehre \citep{nordlund1997stellar}. 

In summary, based on the above results and comparisons done using data from the CO5BOLD numerical simulation, we conclude that 0.63 for \(\Gamma_{1min}\) is also an optimal choice for detecting vortices from numerical simulation data. 

\subsection{Validation with observational data} \label{sst validation}

The data analyzed in this subsection consists of high-resolution photospheric images centered on the Fe I 630.25 nm spectral line, with a spectral window width of 0.45 nm. These observations were acquired using the CRisp Imaging SpectroPolarimeter \citep[CRISP;][]{scharmer2006comments, scharmer2008crisp} on the Swedish 1-meter Solar Telescope \citep[SST;][]{scharmer20031}. Conducted on July 7, 2019, between 08:23:36 UT and 08:39:18 UT, the observations targeted a quiet-Sun region near the central meridian. The field of view (FOV), centered at ($x_c = 0^{\prime\prime}$, $y_c = -300^{\prime\prime}$), covered an area of $56.5^{\prime\prime} \times 57.5^{\prime\prime}$. The pixel size of the data is $0.059^{\prime\prime}$ ($\sim$ 43.6 km), with a spatial resolution estimated to be at least 87.2 km, corresponding to twice the pixel size. The images were taken with an average cadence of 4.2 seconds, and the FOV was rotated 70 degrees clockwise relative to the Sun's north pole. 

\begin{figure*}
   \centering
   \includegraphics[width=1.0\textwidth]{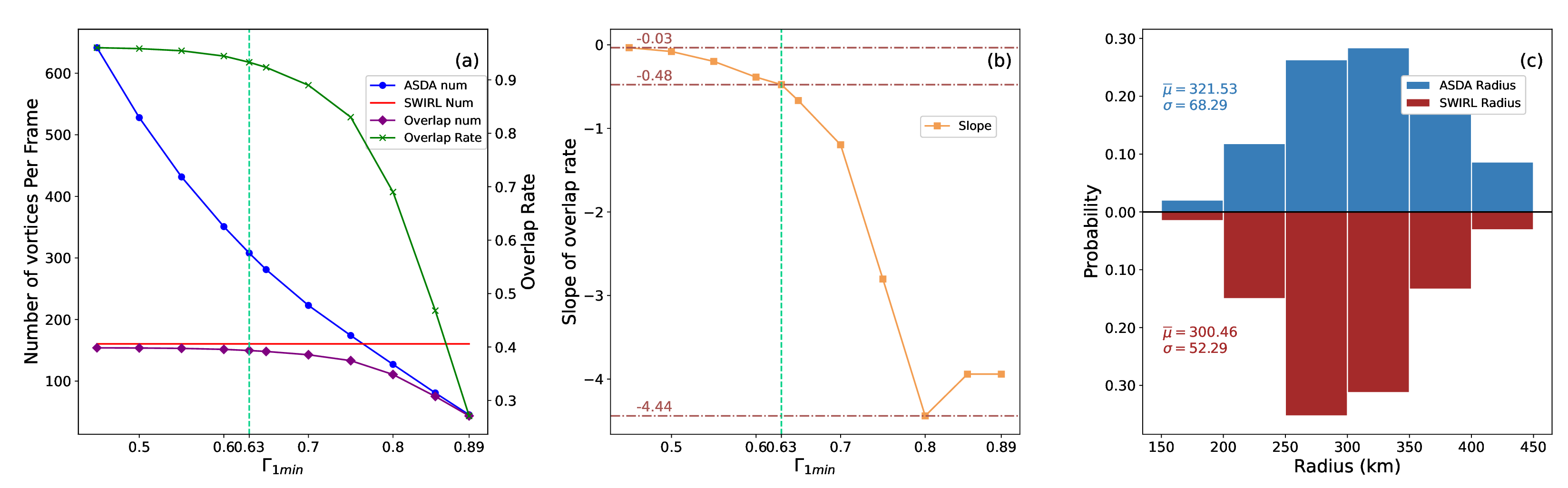}
   \caption{Similar to Fig. \ref{co5bold} but for comparing the detection results obtained from the Optimized ASDA and SWIRL with observational data.}
    \label{sst}%
              
\end{figure*}

{\hl Figure~\ref{vortex_exam_b} also presents an example photospheric intensity map of SST observations in panel (c). Blue and red curves outline the boundaries of vortices detected by ASDA. Correspondingly, panel (d) shows the close-up view of the yellow box in panel (c).}

Similar to Figure~\ref{co5bold}, Figure~\ref{sst} shows the comparison results between vortices detected by ASDA and SWIRL from the SST observations. The number of vortices detected by ASDA and SWIRL is shown with the blue curve and the red horizontal line in Figure~\ref{sst}(a), with the purple and green curves depicting the number of the overlapping vortices and the overlapping rate. Figure~\ref{sst}(b) shows the slope of each point along the green curve in panel (a), with slopes at \(\Gamma_{1min}\) = 0.45, 0.63 and 0.80 as -0.03, -0.48 and -4.44, respectively. The quick drop of the slope from \(\Gamma_{1min}\) = 0.63 to 0.80 also happens, similar to Figure~\ref{co5bold}(b) and (e) for the numerical simulation data. Moreover, the detected radii by Optimized ASDA and SWIRL also show almost the same distributions, with close expected values (322.24 km vs. 300.46 km) and standard deviations (67.54 km vs. 52.29 km), as shown in Figure~\ref{sst}(c). The above results are highly consistent with the vortex detection results from numerical simulation data in Sect.~\ref{simulation}, indicating that Optimized ASDA with \(\Gamma_{1min}\) = 0.63 also performs well in detecting vortices from the solar observational data. 

We have also explored the influence of kernel sizes on the detection of vortices using both numerical simulation data and observational data. We detect vortices from \(\Gamma_1\) and \(\Gamma_2\) calculated with VGCM, VGCM-o, and a single kernel size 7, all with the optimal \(\Gamma_{1min}\) = 0.63. The results are similar, and here, we take the results from the photospheric simulation data for example. There are more vortices ($\sim$ 9\%) detected with VGCM and VGCM-o than with a single kernel size 7 and correspondingly, more vortices detected by SWIRL are overlapped by vortices detected with VGCM and VGCM-o. These results support the results in Sect.~\ref{ks} that the Variable \(\Gamma\) Calculating Method is more suitable than a single kernel size for calculating \(\Gamma_1\) and \(\Gamma_2\). Moreover, the number of vortices detected with VGCM-o is slightly more ($\sim$ 2\%) than the number detected with VGCM, but the numbers of overlapping vortices with SWIRL are identical. This indicates that it is very likely that the additional vortices detected with VGCM-o are false detections, supporting our previous conclusions that VGCM is more accurate in detecting vortices than VGCM-o.

\section{Conclusions and discussions} \label{conclus and discuss}
In this paper, we employed the automated swirl detection algorithm (ASDA, an automated algorithm based on the \(\Gamma\)-functions method) to detect vortices from synthetic data generated with diverse conditions. We also aimed to improve the \(\Gamma\)-functions method for vortex identification. We analysed the effect varying the values of \(\Gamma_{1min}\), which determines the centers of vortices, and applied various ways of calculating \(\Gamma_1\) and \(\Gamma_2\) to search for an optimal value of \(\Gamma_{1min}\) and the best method to calculate \(\Gamma_1\) and \(\Gamma_2\). ASDA with the above improvements is named Optimized ASDA. In this section, we briefly summarize our results and present some discussions on the potential implications of the Optimized ASDA.

In the first stage of this work, we fixed the kernel size $ks$ = 7 to calculate \(\Gamma_1\) and \(\Gamma_2\) and applied ASDA with different values of \(\Gamma_{1min}\) to synthetic data 1 (SD1). No matter the velocity noise is 0 or 20\%, the detection rates showed little difference when \(\Gamma_{1min} \leq\) 0.60 but decreased quickly once \(\Gamma_{1min} > 0.60\). For SD1, \(\kappa\), which was defined as \(\kappa = v_e / v_r\) in Sect. \ref{gammamin}, was set to obey the Gaussian distribution N(0.9, 0.2$^2
$). Theoretically, about 99.7\% vortices could be detected by ASDA with \(\Gamma_{1min}\) equaling 0.55 or less, based on the deductions in the third paragraph in Sect.~\ref{method}, which was proven by the experimental results shown in Tables~\ref{SD1-0} and~\ref{SD1-0.2}. By exploring the variation of the radii of vortices (larger and smaller) in SD1 and repeating the experiments, we found similar results, which indicated that ASDA performs well in detecting vortices with different radii. 

Via changing $\kappa$ to smaller (N(0.5, 0.1$^2$)) and larger (N(1.2, 0.2$^2$)) values, we built two new datasets SD2 and SD3. Similar results were found that when \(\Gamma_{1min}\) is less than 0.60, the detection rate by ASDA is almost invariable. It is worth noting that, the detection rate ($\sim$50\%) of vortices at $\Gamma_{1min}$ = 0.45 on SD3 with a noise level of 20\% is lower than expected (99.7\%). By studying an example region in SD3 with 10 synthetic vortices, we found that some candidates were excluded by the $\Gamma_2$ criterion when $\Gamma_{1min}$ is too small. These results suggest that negative impacts on the performance of ASDA would be introduced with either too-big or too-small values of $\Gamma_{1min}$. Further tests on more synthetic data revealed an optimal value of 0.63 for $\Gamma_{1min}$. 


Next, we fixed the \(\Gamma_{1min}\) to 0.63 and searched for an appropriate method to calculate \(\Gamma_1\) and \(\Gamma_2\). Motivated by the adaptive version of the Advanced $\Gamma$ Method proposed by \cite{yuan2023advanced}, we presented the Variable $\Gamma$ Calculating Method (VGCM) to calculate the two \(\Gamma\) functions. To explore the best method, contrast experiments on SD2 were conducted by using different calculating method: single kernel sizes (ranging from 3 to 15) and several versions of the Variable $\Gamma$ Calculating Method (VGCM-o, VGCM-1, VGCM, and VGCM-2), with results shown in Table \ref{SD1-1.2-0.2-VGM-o}. False vortices were detected only when using a single kernel size $ks$ = 3 and VGCM-o (kernel sizes = 3, 5, 7, 9, and 11), which indicated $ks$ = 3 resulted in poor detection results. Moreover, by comparing the detection results with other methods, shown in Table \ref{SD1-1.2-0.2-VGM-o}, we found that the Variable \(\Gamma\) Calculating Method performed better than single kernel size, and VGCM (with kernel sizes = 5, 7, 9, and 11) costed less computing sources than VGCM-1 and VGCM-2 but still revealed similar performance in detecting vortices in SD2. Similar comparison results were obtained when changing the radii of vortices in SD2 to larger and smaller. These results suggest VGCM is the most suitable method to calculate \(\Gamma_1\) and \(\Gamma_2\).

After finding 0.63 is the optimal value of \(\Gamma_{1min}\) and VGCM (kernel sizes = 5, 7, 9, and 11) is more appropriate for calculating \(\Gamma_1\) and \(\Gamma_2\), ASDA can be optimized for more accurate vortex identification, named as the Optimized ASDA. To validate the reliability of the Optimized ASDA, we employed it to detect small-scale vortices in numerical simulation data of the solar atmosphere from the radiative MHD CO5BOLD code and observational data of the photosphere by SST. The comparison results are all similar and consistent with the conclusions in Sect.~\ref{gammamin} and Sect.~\ref{ks} that the choice 0.63 of \(\Gamma_{1min}\) and the application of VGCM to calculate \(\Gamma_1\) and \(\Gamma_2\) are both more suitable than the original ASDA. However, we noted that the numbers of vortices detected by the Optimized ASDA were all more than the numbers detected by SWIRL, showing 39.8\%, 80\%, and 91.3\% more vortices for the photospheric, chromospheric simulations, and the SST photospheric observations, respectively (see Fig.~\ref{co5bold}a, d and Fig.~\ref{sst}a). A possible reason is that SWIRL missed some vortices. \cite{cuissa2022innovative} and \cite{cuissa2024innovative} noted two drawbacks of SWIRL: (1) the detection is not strictly Galilean invariant, which means some vortices with rotation speeds comparable to the flow speeds could be missed by SWIRL. They also pointed out that this shortcoming should not affect photospheric vortices because they are predominantly rooted in intergranular lanes and moving slowly relative to the vortical flow speed \citep{tziotziou2023vortex}. (2) the parameters for clustering and detection in SWIRL call for adjustments when applied to different data. 
\begin{figure}
   \centering
   \includegraphics[width=0.5\textwidth]{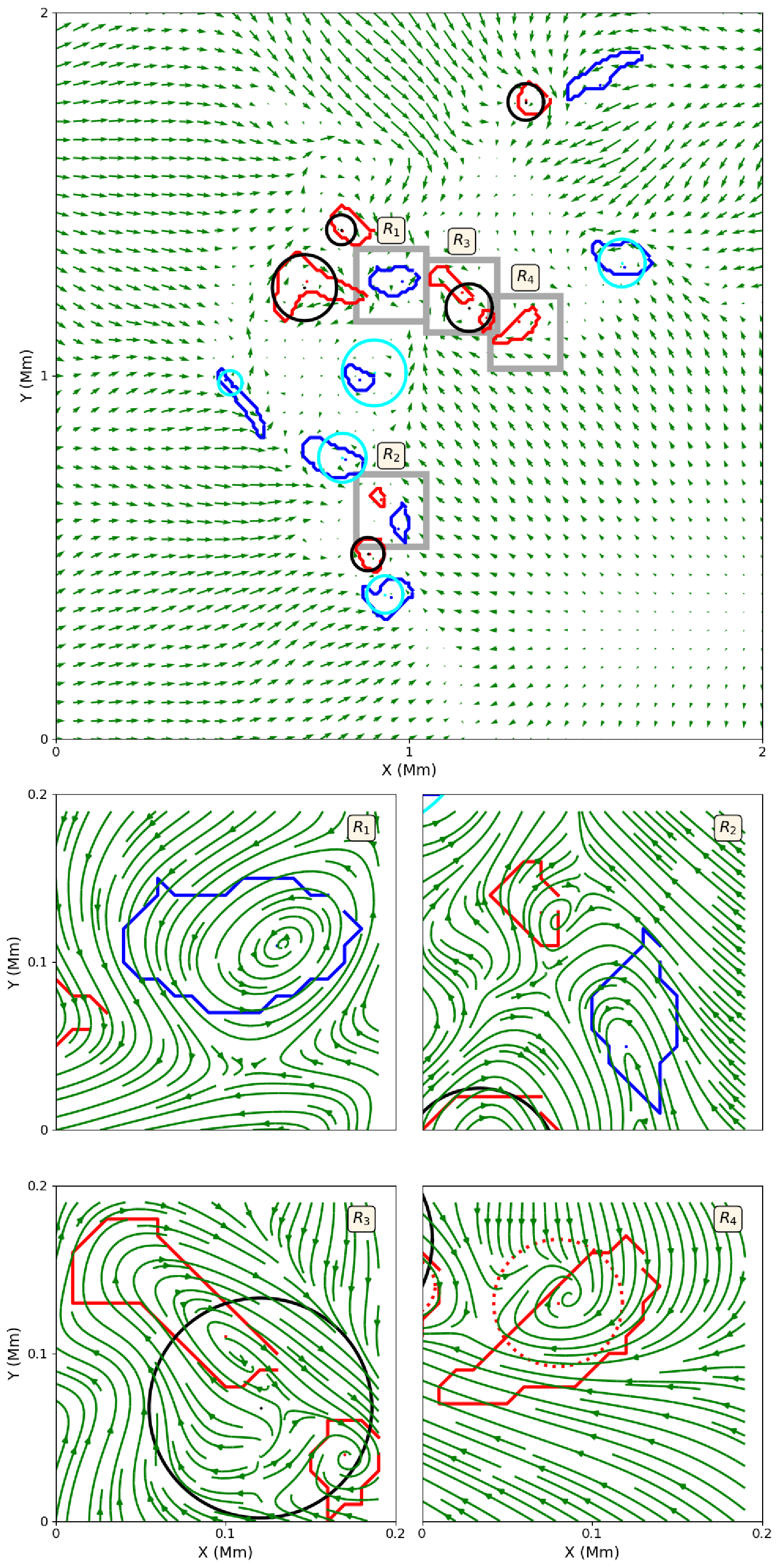}
   \caption{\textbf{Top panel}: A small region of the CO5BOLD photospheric simulation domain. Green arrows represent the velocity field. Blue and cyan colors are vortices detected by the Optimized ASDA and SWIRL rotating counter-clockwise. In contrast, the red and black ones rotate clockwise for swirls detected by the Optimized ASDA and SWIRL, respectively. \textbf{Bottom panel}: Close-up views of four rectangle regions outlined with grey {\hl squares} in the top panel.}
    \label{compare}%
              
\end{figure}

The above two reasons could result in the underestimation of the vortices number detected by SWIRL, but whether the additional vortices detected by the Optimized ASDA in Figure~\ref{co5bold}(a), (d) and Figure~\ref{sst}(a) are true vortices or not needs further exploration. {\hl In other words, the optimized ASDA might have overestimated the number of vortices, which could also contribute to the fact that there have been more vortices detected by the Optimized ASDA.} 

The top panel of Figure~\ref{compare} shows a small 2 \(\times\) 2 Mm$^2$ region of the photospheric numerical simulation domain in CO5BOLD. The green arrows represent the velocity field, and the blue and red curves show the boundaries of positive (counterclockwise) and negative (clockwise) vortices detected by the Optimized ASDA, while the cyan and black curves depict the boundaries of positive and negative vortices detected by SWIRL. Most vortices are identified by the Optimized ASDA and SWIRL at the same time, and the effective radii decided by the Optimized ASDA and SWIRL are also similar, which is consistent with the similar vortex radius distributions in Figure~\ref{co5bold}(c). However, some vortices are only detected by  the Optimized ASDA, shown in the regions labeled as $R_1$, $R_2$, and $R_4$ outlined with grey {\hl squares}. The middle and bottom panels of Figure~\ref{compare} depict close-up views of the three regions. In panel $R_1$, the positive vortex identified by Optimized ASDA appeared to be an actual vortex, but SWIRL missed it. Similar circumstances also occurred in panels \(R_2\) and \(R_4\). It suggests that SWIRL does ignore some true vortices detected by Optimized ASDA, explaining the number gaps in Figure~\ref{co5bold}(a), (d), and Figure~\ref{sst}(a). {\hl On the other hand, it is somewhat intriguing that the counter-clockwise vortex in $R_2$ seems to be not true according to the streamline plot, while the other clockwise one is really a true one. This result further supports the concerns raised earlier that the Optimized ASDA could detect some false vortices.}

It is observed that the boundary of the vortex detected by the Optimized ASDA in panel \(R_4\) does not quite conform to the velocity field {\hl (the red dotted circle seems more suitable)}. It suggests that the algorithm for determining the vortex boundaries by the Optimized ASDA remains to be improved. Moreover, we note that a vortex identified by SWIRL (outlined by the black circle in Fig.~\ref{compare} $R3$) is detected as two separate vortices by the Optimized ASDA. However, investigating the velocity field, we can see that the vortex identified by SWIRL is not a real one. The oval vortex at the top left and another at the bottom right, detected by the Optimized ASDA, are, in turn, more consistent with the actual velocity field. These observations suggest that the Optimized ASDA performs better than SWIRL in detecting non-standard shaped vortices and yielding a more accurate number of vortices in the solar atmosphere. {\hl It is worth noting that, in this work, noise is inserted into the velocity map to generate the synthetic data. However, whether this kind of synthetic data is well-suited for generating non-standard vortices (like vortices detected by the Optimized ASDA in Fig.~\ref{compare} R3) remains unclear. One future task conducting a more detailed analysis of non-standard vortices could lead to further improvement of ASDA and other vortex identification methods.} 

\cite{liu2019evidence} found that abundant photospheric vortices excite Alfv{\'e}n pulses, which propagate upward and carry energy flux into the upper chromosphere. They noted that the energy flux (\(F_A\)) carried into the upper chromosphere by a single Alfv{\'e}n pulse is estimated to be 1.9-7.7 kW m$^{-2}$. The average energy flux (\(\overline{F_A}\)) is defined as:
\begin{equation}
    \overline{F_A}=\frac{F_A\overline{N}\pi\overline{R}^2}{S_{FOV}},
    \label{energy flux}
\end{equation}
where \(\overline{N}\) and \(\overline{R}\) are the average number of vortices in each frame and vortex effective radius. \(S_{FOV}\) represents the area of the field-of-view (FOV) of the observation. We employed the original ASDA with \(\Gamma_{1min}\) = 0.89 and the Optimized ASDA with \(\Gamma_{1min}\) = 0.63 to the SST observation mentioned in Sect.~\ref{sst validation}. On average, 39.6 and 308 vortices are detected by the original ASDA and the Optimized ASDA, with corresponding average vortex radius of 308 km and 271 km, respectively. Therefore, using Eq. (\ref{energy flux}), \(\overline{F_A}\) is found to be around 12.6-51.2 W m$^{-2}$ and 75.9-308.3 W m$^{-2}$ by employing the original ASDA and the Optimized ASDA, respectively. The former flux is not enough to balance the local radiative energy losses (\(\sim\) 100 W m$^{-2}$) \citep{withbroe1977mass} in quiet-Sun regions. 
{\hl It is worth noting that, as mentioned before, the Optimized ASDA might overestimate the number of vortices, and SWIRL could underestimate the number. Thus, the energy flux (75.9-308.3 W m$^{-2}$) estimated from the Optimized ASDA can be viewed as an upper limit of the flux supplied by the photospheric vortices. Meanwhile, we can also provide a lower limit, which is around 39.4-160.2 W m$^{-2}$, by using the number (160) of vortices identified by SWIRL. The above results from the Optimized ASDA and SWIRL indicate that the average energy flux related to photospheric vortices is very likely enough to balance the energy losses.}
It further supports the fact that prevalent photospheric vortices could play significant roles in heating the upper atmosphere \citep[e.g.,][]{shelyag2013alfven, chmielewski2014numerical, mumford2015generation, mumford2015photospheric, liu2019evidence, battaglia2021alfvenic}. 

\(\Gamma\)-functions method (and relative automated algorithms, such as ASDA) for vortex identification heavily depends on the estimated horizontal velocity field. Methods used to calculate the horizontal velocity fields all have their drawbacks. For example, the most common technique FLCT we used in this work should be applied with caution when estimating granular and subgranular flows \citep{tremblay2018reconstruction, cuissa2024innovative}. Moreover, \cite{verma2013evaluating} and \cite{liu2019automated, liu2019evidence} pointed out that FLCT underestimates the horizontal velocity field by a factor of approximately three and influences the characteristics of detected vortices, such as the rotation speed and expansion speed. In this work, synthetic data was employed to improve the \(\Gamma\)-functions method. Therefore, our results are general and independent of the velocity estimation method. 

Concerning the significant influence of the reconstructed velocity fields, a key point of future work is to check the reliability of different velocity estimation methods and search more reliable approaches for different observations. 

One of our recent studies \citep{Liu2025suvel} used a neural network technique trained on high-resolution data \citep[with a pixel size of $\sim$12 km, comparable to the diffraction limit of the Daniel K. Inouye Solar Telescope, DKIST,][]{Rimmele2020} from realistic radiative numerical simulations of the solar photosphere. The built neural network model performs significantly better than FLCT at these small scales. It is worth investigating how these different methods of estimating the photospheric horizontal velocity fields would affect the vortex detection results by the Optimized ASDA.

\begin{acknowledgements}
We thank Dr. Fabio Riva in Istituto ricerche solari Aldo e Cele Daccò (IRSOL) for the source of the CO5BOLD simulation data. We acknowledge the support from the Strategic Priority Research Program of the Chinese Academy of Science (Grant No. XDB0560000) and the National Natural Science Foundation (NSFC 42188101, 12373056). R.E. is grateful to Science and Technology Facilities Council (STFC, grant No. ST/M000826/1) UK, acknowledges NKFIH (OTKA, grant No. K142987 and Excellence Grant, grant nr TKP2021-NKTA-64) Hungary and PIFI (China, grant number No. 2024PVA0043) for enabling this research. This work was also supported by the International Space Science Institute project (ISSI-BJ ID 24-604) on "Small-scale eruptions in the Sun". 

\end{acknowledgements}

%
%
\bibliographystyle{aa}
\bibliography{ref}

\label{LastPage}
\end{document}